\documentclass[a4paper,11pt]{article}
\pdfoutput=1 

\usepackage{jheppub} 

\usepackage[T1]{fontenc} 

\usepackage{color}

\title{\boldmath Revisiting constraints on 3+1 active-sterile neutrino mixing using IceCube data}

\author{Luis Salvador Miranda,}
\author{Soebur Razzaque}
\affiliation{Centre for Astro-Particle Physics (CAPP) and Department of Physics,\\ University of Johannesburg, PO Box 524, Auckland Park 2006, South Africa}

\emailAdd{smiranda-palacios@uj.ac.za}
\emailAdd{srazzaque@uj.ac.za}

\abstract{Recent IceCube search results for sterile neutrino increased tension between the combined appearance and disappearance experiments. On the other hand, MiniBooNE latest data confirms at $4.9\sigma$ CL the short-baseline oscillation anomaly. We analyze published IceCube data based on two different active-sterile mixing schemes using one additional sterile neutrino flavor. We present exclusion regions in the parameter ranges $0.01 \le \sin^2 \theta_{24} \le 0.1$ and $0.1~{\rm eV}^2 \le \Delta m^2_{42} \le 10~{\rm eV}^2$ for the mass-mixing and flavor-mixing schemes. Under the more conservative mass-mixing scheme, $3\sigma$ CL allowed regions for the appearance experiment and MiniBooNE latest result are excluded at $\gtrsim 3\sigma$ CL.  In case of less-restrictive flavor-mixing scheme, results from the appearance experiments are excluded at $\gtrsim 2\sigma$ CL.  We also find that including prompt component of the atmospheric neutrino flux relaxes constraints on sterile mixing for $\Delta m^2_{42} \gtrsim  1~{\rm eV}^2$. }

\arxivnumber{0}

\begin{document} 
\maketitle
\flushbottom

\section{Introduction}
\label{sec:intro}

The anomalies reported by the LSND~\cite{Aguilar:2001ty} and MiniBooNE~\cite{AguilarArevalo:2010wv} detectors (Short Baseline experiments); reactor~\cite{Mention:2011rk, Mueller:2011nm} and Gallium experiments~\cite{Abdurashitov:2005tb}, can not be explained within the standard 3-$\nu$ oscillation framework. A possible solution is to consider the existence of additional neutrino flavors, which are sterile with respect to electroweak interactions. In fact, addition of one sterile neutrino of mass $\sim 1$~eV fits the data well. Sterile neutrino mixing with active neutrinos have been studied for the last decades as an extension of the Standard Model, to study and explain these anomalies~\cite{Babu:2016fdt, Giunti:2017yid, Gariazzo:2017fdh, Canas:2017umu,Bertuzzo:2018itn, Ballett:2018ynz,Rajpoot:2013dha}.  On the other hand, constraints have been derived on the sterile neutrino parameters using accelerator data that do not involve oscillations~\cite{Das:2017zjc, Kim:2018uht}. One also needs to consider that as extra neutrinos are more massive, it is difficult to fit within the cosmological constraints. In fact recent cosmological data disfavor a heavy sterile neutrino~\cite{Izotov:2010ca, Steigman:2010pa, Giusarma:2011ex, Forastieri:2017oma, Feng:2017nss}, although other recent works propose alternative solutions to this conflict~\cite{Chauhan:2018dkd, Song:2018zyl, Berryman:2018jxt, Bezrukov:2017ike, Archidiacono:2016kkh}. 

Several experiments have been conducted recently in search of sterile neutrinos~\cite{Ko:2016owz, Adamson:2016jku, MINOS:2016viw, Adamson:2017uda, Agafonova:2018dkb, Aartsen:2017bap}, while many phenomenological studies have been performed to shed lights on anomalies in the reactor and Gallium experiments~\cite{Gariazzo:2018mwd, Barinov:2017ymq, Dentler:2017tkw}, as well as on the current and future short- and long-baseline experiments~\cite{Thakore:2018lgn, perron, Capozzi:2016vac}. Recent IceCube and MINOS results show a strong tension between the appearance and disappearance experiments. IceCube collaboration reported a search for the $\nu_{\mu}+\bar{\nu}_{\mu}$ disappearance results for the IC86 string configuration data taken during the 2011-2012 period; which excluded the allowed region of the appearance experiments, including the LSND and MiniBooNE, at $99\%$~CL~\cite{TheIceCube:2016oqi}. Further studies have been performed to shed light on IceCube results~\cite{Liao:2016reh, Esmaili:2018qzu, Petcov:2016iiu, Brdar:2016thq, Collin:2016aqd, Moss:2017pur}. However, the latest MiniBooNE results combined with LSND results amount to a \textcolor{black}{$6.0\sigma$} evidence for new physics beyond the Standard Model~\cite{Aguilar-Arevalo:2018gpe}.

Propagation of sterile neutrino over long baseline through the Earth's mantle and/or core has an effect on muon neutrino oscillations due to matter effect and MSW resonance \cite{Nunokawa:2003ep, Choubey:2007ji, Razzaque:2011ab, Esmaili:2012nz, Esmaili:2013vza, Lindner:2015iaa}, which produces an enhancement to the $\nu_{\mu}-\nu_{s}$ oscillations, thus causing a depletion in the muon neutrino flux, with a distortion of the energy and zenith angle distributions of atmospheric neutrino events in a detector. IceCube analyses take into account conventional atmospheric neutrino flux  contribution~\cite{TheIceCube:2016oqi}, but for the range of neutrino energy up to $10^6$ GeV the prompt atmospheric neutrino flux contribution could change the total number of events measured by the detector. Furthermore, different mixing schemes of sterile neutrinos with active neutrinos can change the exclusion regions derived from the IceCube data.

In this paper we analyze IceCube public data in order to derive constraints on mixing angle and mass-square difference with active neutrinos. We use the mass- and flavor-mixing schemes for a sterile neutrino in 3+1 scenario~\cite{Razzaque:2011ab} as well as prompt atmospheric neutrino flux~\cite{Enberg:2008te} contributions to the data. The paper is divided into the following sections: In Section 2 we explain probabilities of the mass-mixing and $\nu_{s} - \nu_{\mu}$ flavor-mixing schemes. In Section 3 we use conventional atmospheric neutrino flux and 
prompt contribution to calculate muon neutrino flux at the IceCube detector. In Section 4 we obtain the number of events based on the IceCube detector data tensors and fluxes of the last section. In Section 5 we use a $\chi^2$ statistical test for studying the models and derive constraints on the mixing parameters. Finally, in section 6 we give results and summary.

\section {Sterile mixing schemes and oscillation probabilities}
The 3+1 active-sterile neutrino general mixing scheme consists of 4 flavor states $\nu_{f}^T = (\nu_{e}, \nu_{\mu}, \nu_{\tau}, \nu_{s})$ and 4 mass states $\nu_{\rm mass}^T = (\nu_{1}, \nu_{2}, \nu_{3}, \nu_{4})$ with relation $\nu_{f} = U_{f}\nu_{\rm mass}$. The unitary matrix $U_{f}$ takes the form $U_{f} = R_{34}R_{24}R_{14}R_{23}R_{13}R_{12}$, where $R_{ij}$ is a rotation matrix in the $i$--$j$ plane (for matrices with indices $j \neq i + 1$, $\pm \sin\theta_{ij} \longrightarrow \pm \sin\theta_{ij} e^{\mp i \delta_{ij}}$, with CP-violation phase $\delta_{ij}$). At high energies, $E \geq 100$~GeV, electron neutrino effect on muon neutrino probabilities can be neglected due to two factors: first, atmospheric flux of electron neutrinos is very low compared to muon neutrinos; and second, the $\nu_{e}$ is mostly converted to $\nu_{s}$. The CP phase value is also not important. Therefore a good approximation is to neglect the mixing of $\nu_{1}$~\cite{Razzaque:2011ab}. In this approximation the unitary matrix takes the form
\begin{align}
U_{f}=R_{34}R_{24}R_{23}.
\label{eq:1}
\end{align}
Under this parametrization we briefly describe the following two mixing schemes~\cite{Razzaque:2011ab}.

\subsection{Mass mixing scheme}
In this scheme the sterile neutrino mixes with a linear combination of the neutrino mass states 3 and 4, and for this condition the model is also called the  maximal 3-4 mixing scheme. The unitary matrix in this case is parametrized as $U_{f}=U_{23}U_{\alpha}$, where $U_{\alpha}$ is rotation of the neutrino mass states $\nu_{3}$ and $\nu_{4}$ on the angle $\alpha$, and $U_{23}$ is the rotation matrix in 2-3 plane. Therefore,
\begin{align}
U_{f}=\left(
\begin{array}{ccc}
\cos \theta_{23} & -\sin \theta_{23} \cos \alpha & \sin \theta_{23} \sin \alpha \\
\sin \theta_{23}  & \cos \theta_{23} \cos \alpha & -\cos \theta_{23} \sin \alpha \\
0 & \sin \alpha & \cos \alpha
\end{array}
\right).
\end{align}
For values $\sin^2 \theta _{23}= 1/2$, the elements  $U_{\mu 4}$ and $U_{\tau 4}$ are equal. The relation between $\alpha$ and general rotation angles of eq.~(\ref{eq:1}) satisfies the conditions:
\begin{align}
\cos\alpha =\cos\theta_{24}\cos\theta_{34} \,\,,\,\,
\sin\theta_{24} =\sin\theta_{34}/\cos\theta_{34}.
\label{eq:22}
\end{align}

\subsection{$\nu_{s}-\nu_{\mu}$ flavor mixing scheme}
For this scheme, there is no mixing between the 3-4 states ($\theta_{34}=0$) and $U_{f} = U_{24}U_{23}$.  The mixing matrix takes the form
\begin{align}
U_{f}=\left(
\begin{array}{ccc}
\cos \theta_{24}\cos \theta_{23} & -\cos \theta_{24}\sin \theta_{23} & -\sin \theta_{24} \\
\sin \theta_{23} & \cos \theta_{23} & 0 \\
\sin \theta_{24}\cos \theta_{23} & -\sin \theta_{23}\sin \theta_{24} & \cos \theta_{24}
\end{array}
\right).
\end{align}
Therefore, the $\nu_{s}$ state does not mix with $\nu_{4}$ ($U_{\tau 4}=0$).  The LSND and MiniBooNE $\nu_\mu$ oscillation results in this scheme is encoded in the element $U_{\mu 4} = -\sin\theta_{24}$ with mixing angle $\theta_{24}$.  On the other hand, in mass-mixing scheme, the important parameter is $U_{\mu 4} = \sin\theta_{23}\sin\alpha$ and the mixing is governed by the angle $\alpha$. The relationship between $\theta_{24}$ and $\alpha$ in case of $\sin^2\theta_{23} = 1/2$ is the following
\begin{align}
\sin^2\theta_{24} = \frac{\sin^2\alpha}{2-\sin^2\alpha}\,.
\label{alpha_t24}
\end{align}

\subsection{Probabilities}
To calculate the probabilities, we solve numerically the Schr\"odinger equation for neutrino propagation inside the earth in flavor base with the Hamiltonian:
\begin{align}
H_{f}=\frac{1}{2E}U_{f}MU_{f}^{T}+V_{f} 
\end{align}
and with diagonal matrix $M=\textrm{diag}(m_{2}^2,\, m_{3}^2,\, m_{4}^2)$. Here the potential matrix in the flavor base is \textcolor{black}{$V_{f}=\textrm{diag}(0,\, 0,\,-V_{\mu})$}, where we have subtracted matrix $V_{\mu}I$ proportional to the identity matrix $I$.  We also assume normal mass hierarchy and use $\Delta m^2_{32} =2.5 \times 10^{-3}$~eV$^2$ and $\sin^2 \theta _{23}= 1/2$, which are consistent with current experimental results~\cite{Abe:2017vif, NOvA:2018gge, Aartsen:2017nmd}. The matter potential can be approximated as: 
\begin{align}
V_{\mu}=-\frac{G_{\rm F} \rho}{2 \sqrt{2}m_{N}} \approx -1.78\times \textcolor{black}{10}^{-14} 
\left( \frac{\rho}{\rm g/cc} \right)~{\rm eV},
\end{align}
where $\rho$ is matter density calculated from the density profile of the earth according to the Preliminary Reference Earth Model~\cite{Preliminar1}. The probability relation for muon neutrinos between the schemes is given by~\cite{Razzaque:2011ab}
\begin{align}
P^{(f)}_{\mu\mu}=\left( 2 \sqrt{P_{\mu\mu}^{\rm (mass)}}-1 \right)^2 \,,
\end{align}
where $P^{(f)}_{\mu\mu}$ and $P_{\mu\mu}^{\rm (mass)}$ refer to the flavor- and mass-mixing schemes, respectively. For a detailed review of probability calculations please consult reference~\cite{Razzaque:2011ab}.

\begin{figure}[tbp]
\centering
\includegraphics[width=.45\textwidth]{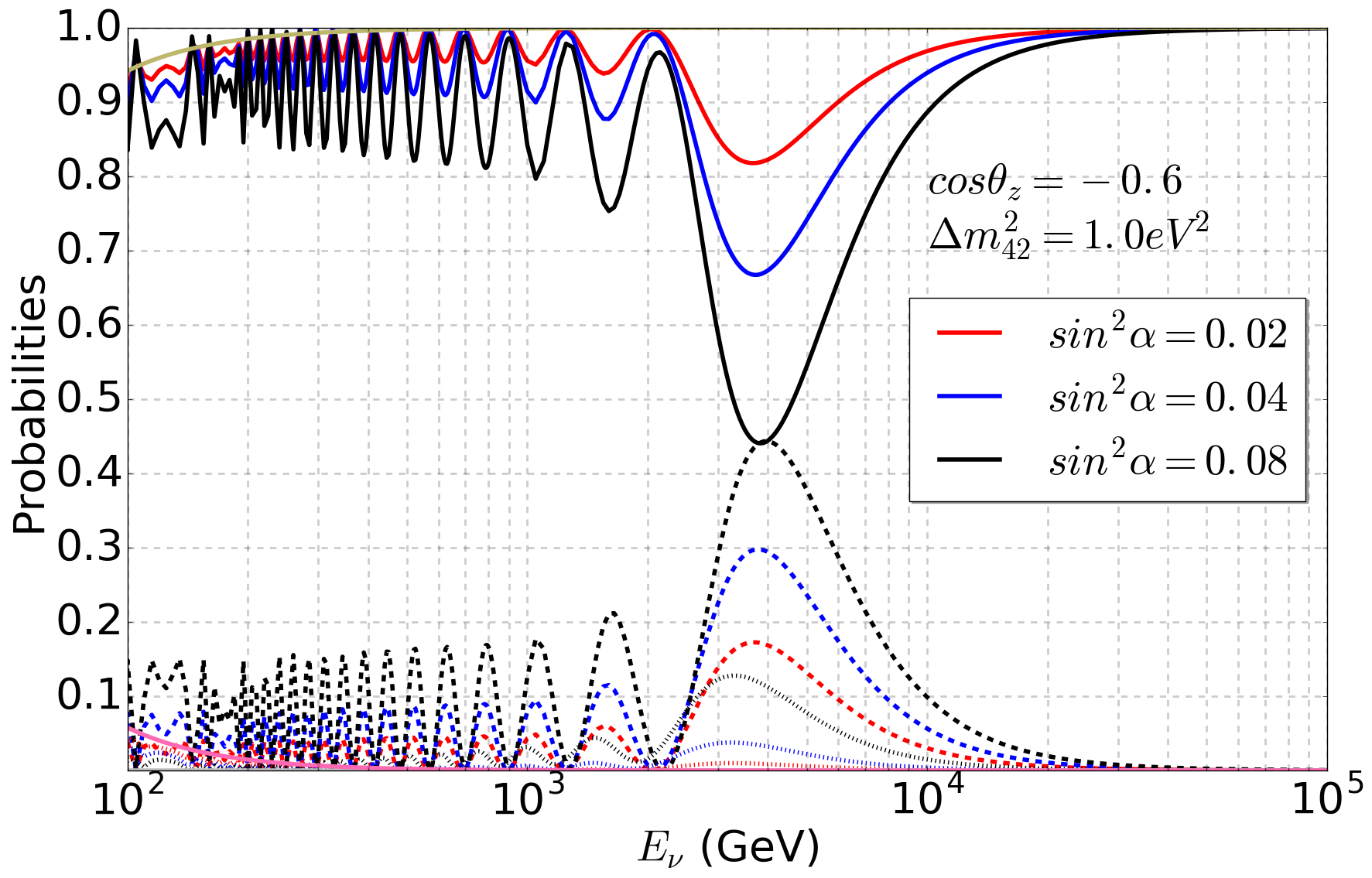}
\includegraphics[width=.45\textwidth]{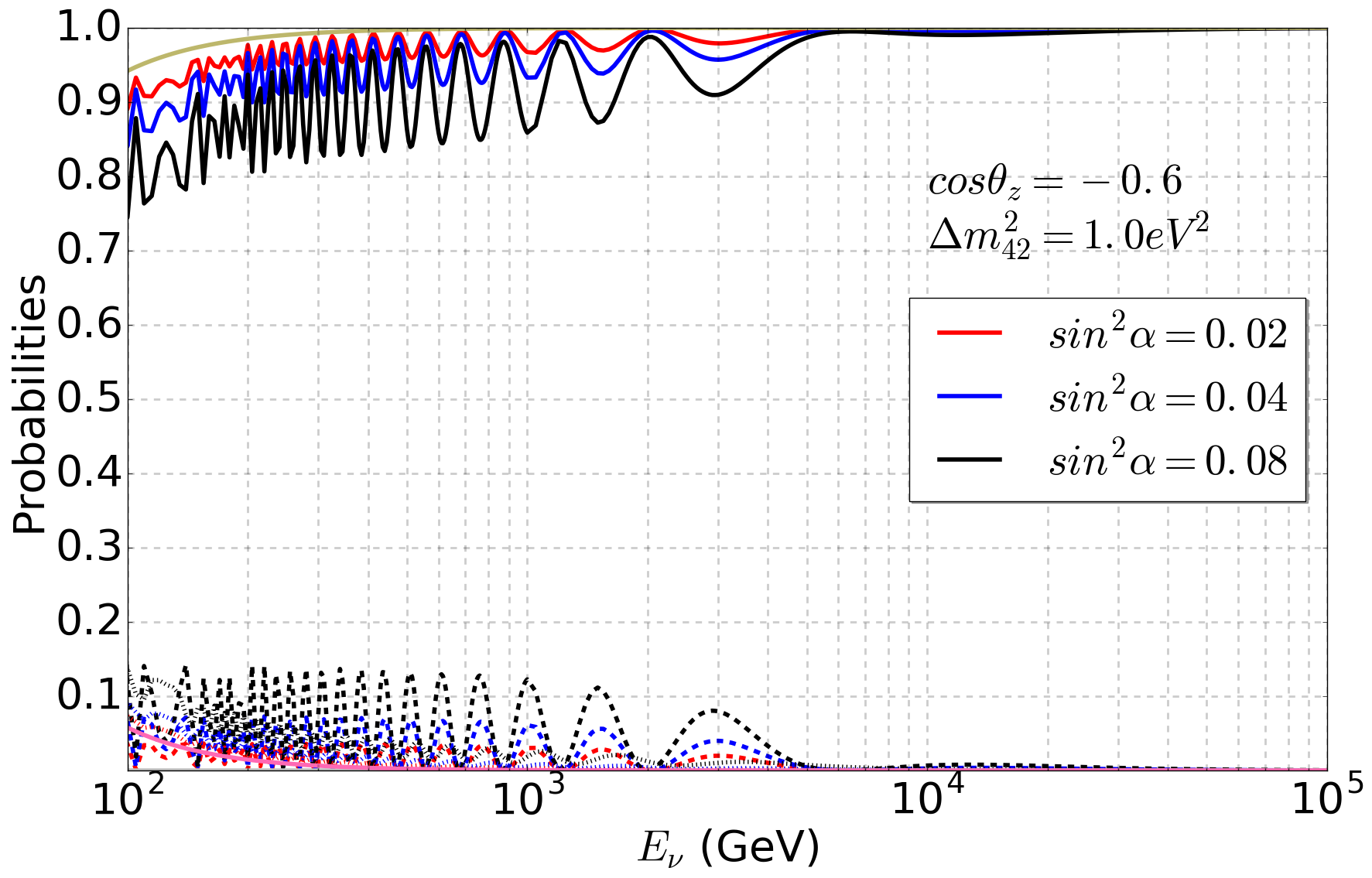}
\includegraphics[width=.45\textwidth]{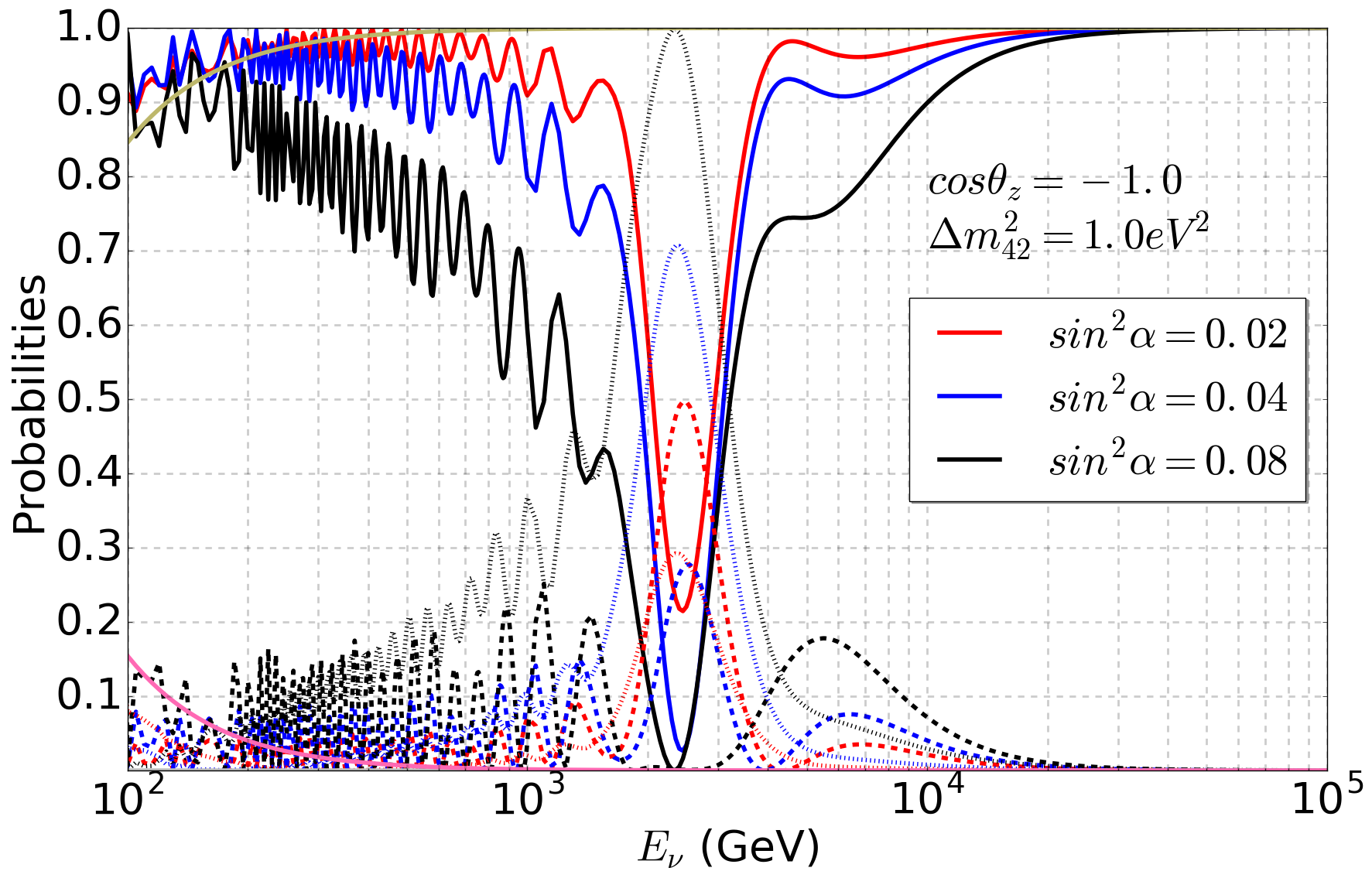}
\includegraphics[width=.45\textwidth]{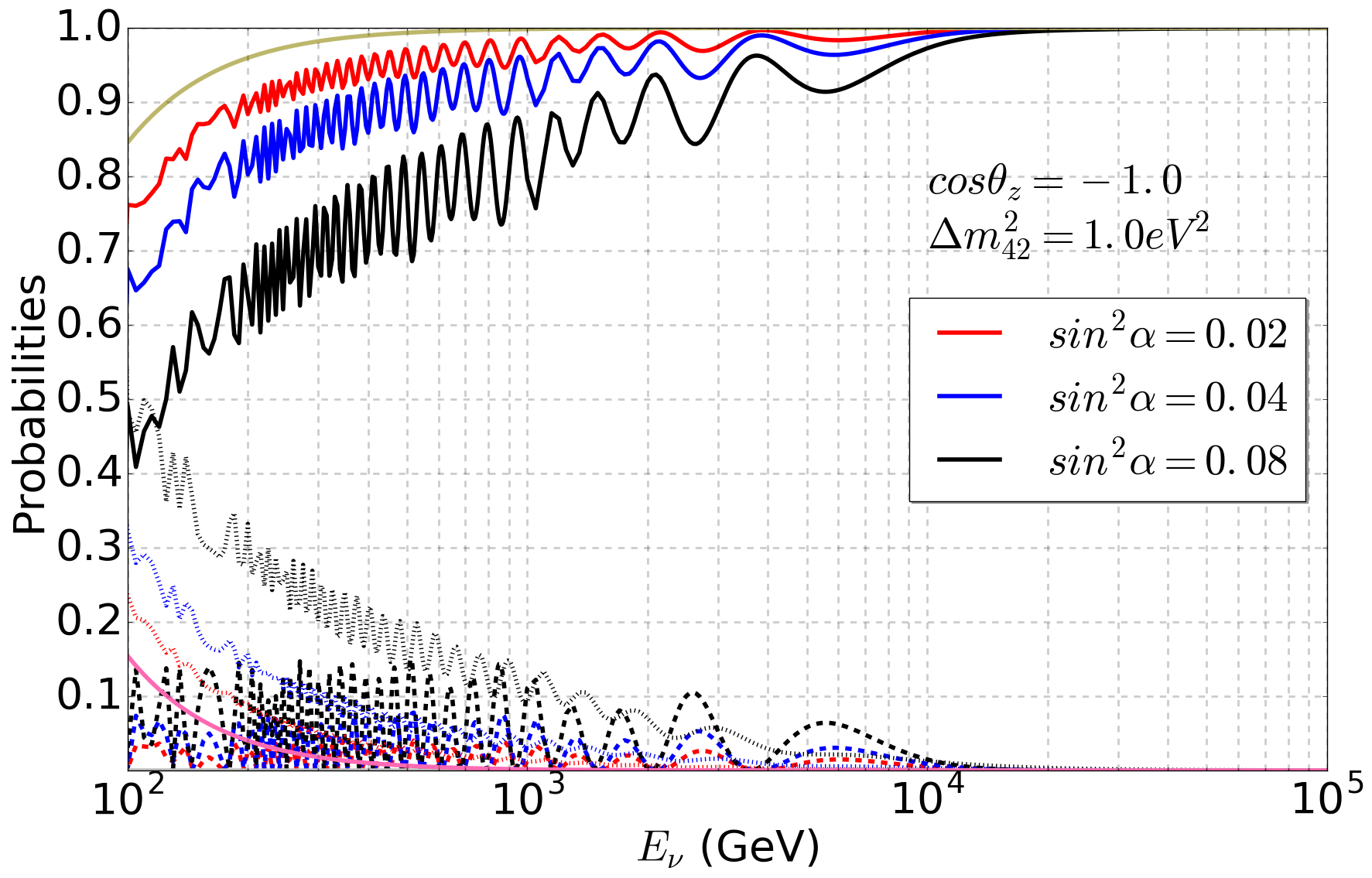}
\caption{\label{fig:prob_mass} Mass mixing scheme probabilities for ${\bar \nu}_\mu$ (left panels) and $\nu_\mu$ (right panels) for a mantle crossing trajectory (top panels) and a mantle-core-mantle crossing trajectory (bottom panels). The continuous lines correspond to $P_{\mu \mu}$, dashed lines to $P_{\mu s}$ and dotted lines to $P_{\mu \tau}$. The green and pink lines (unlabelled) correspond to $P_{\mu \mu}$ and $P_{\mu \tau}$ with no sterile neutrino mixing.}
\end{figure}

\begin{figure}[tbp]
\centering
\includegraphics[width=.45\textwidth]{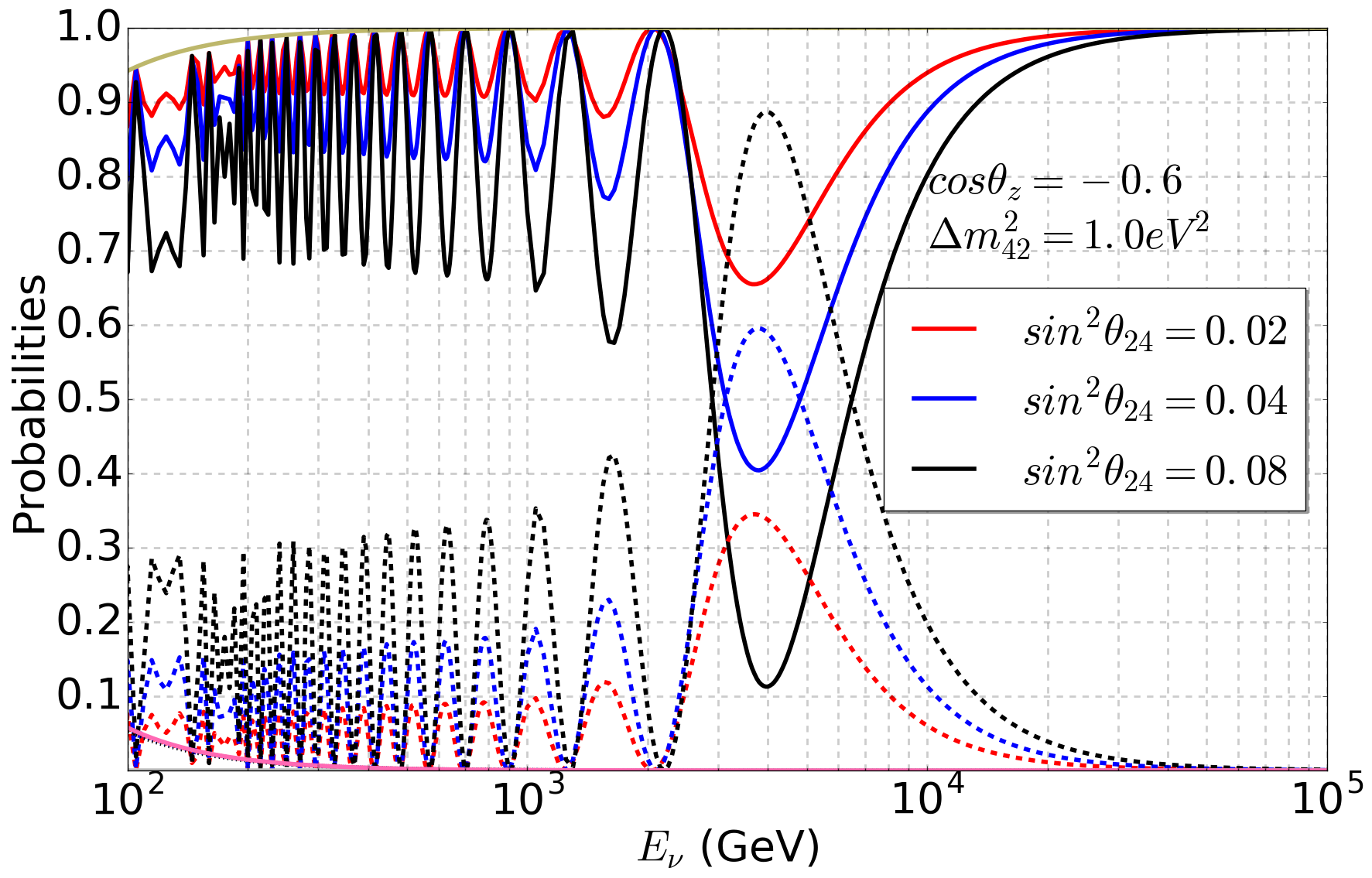}
\includegraphics[width=.45\textwidth]{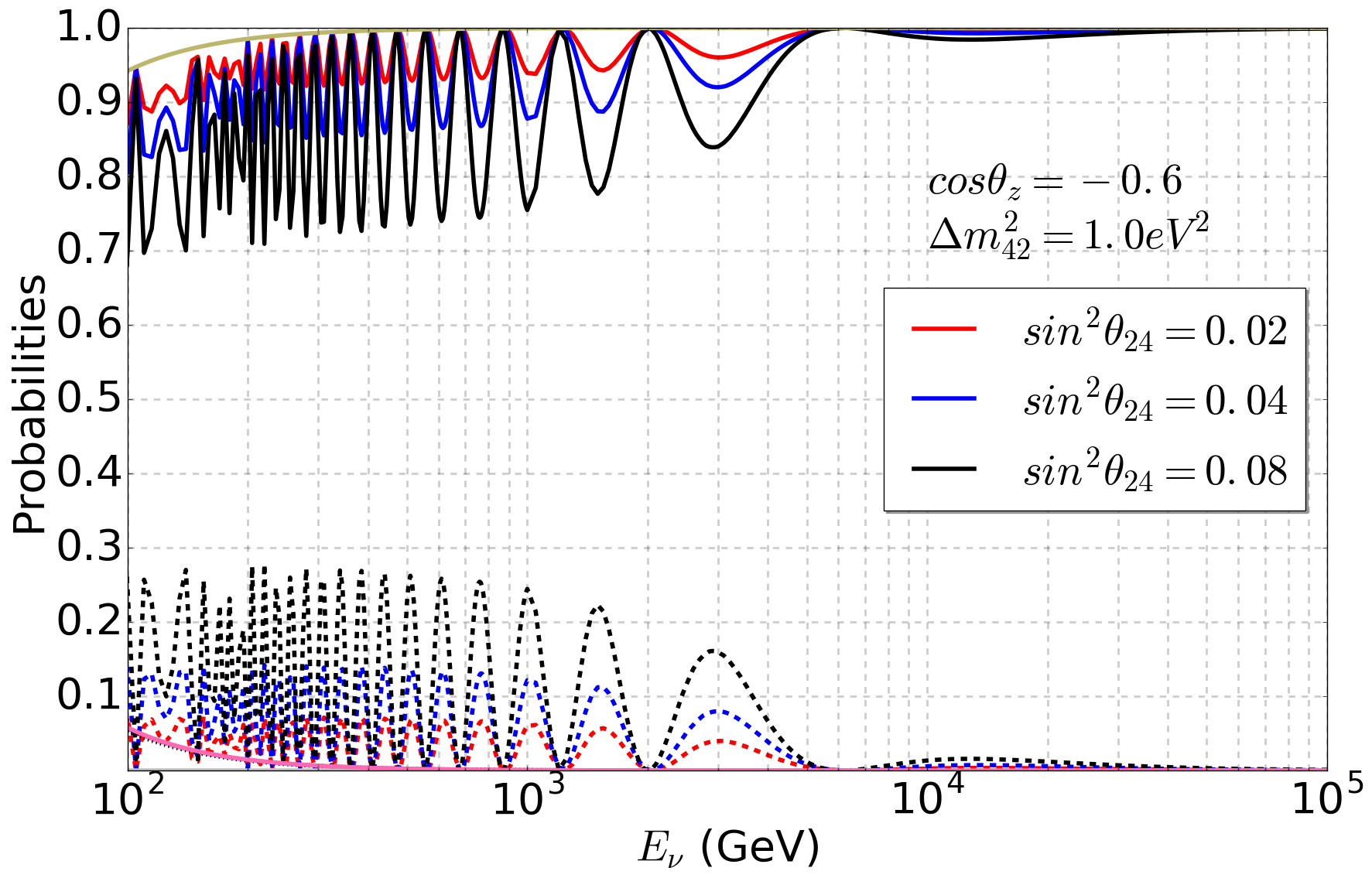}
\includegraphics[width=.45\textwidth]{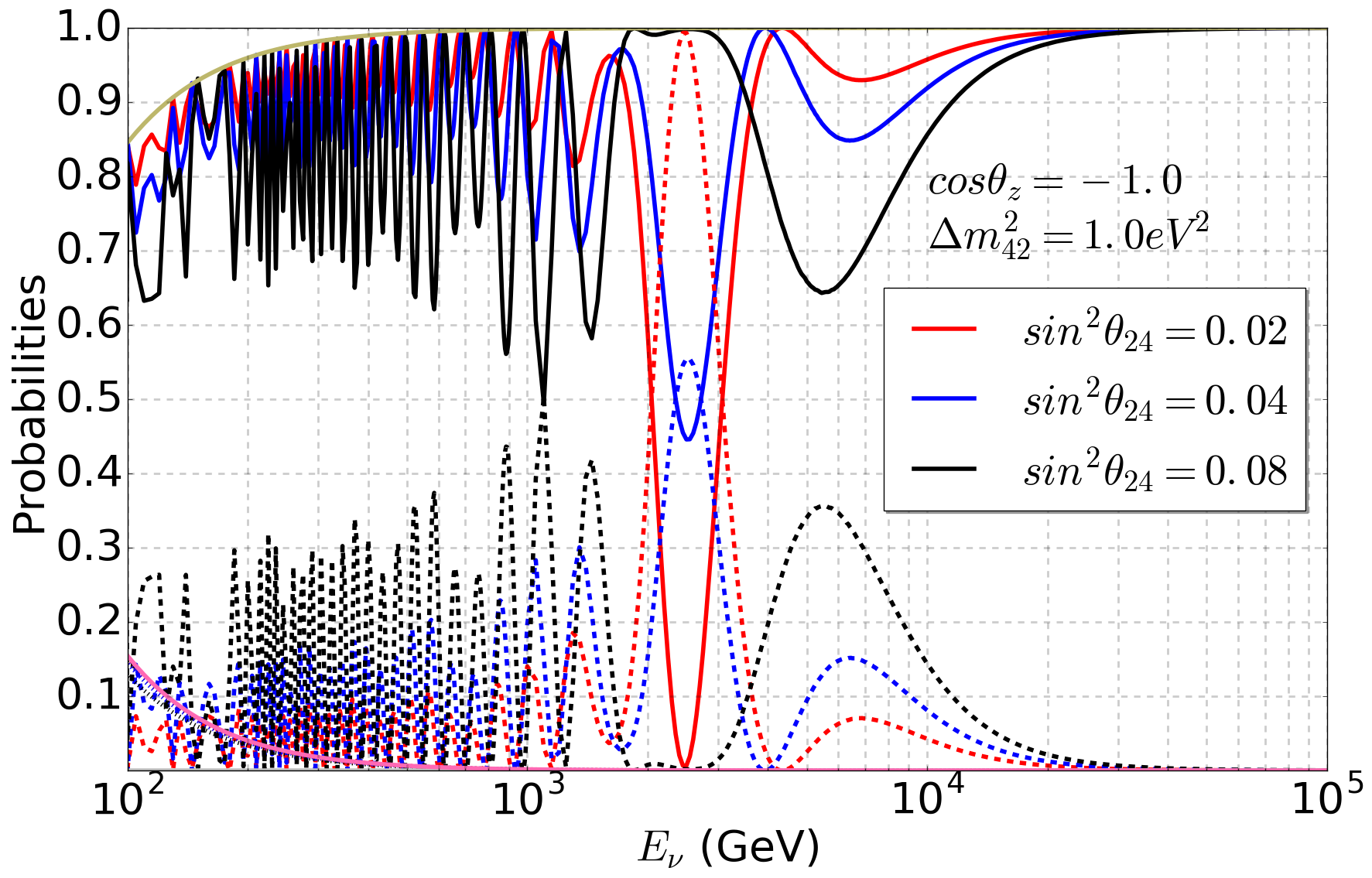}
\includegraphics[width=.45\textwidth]{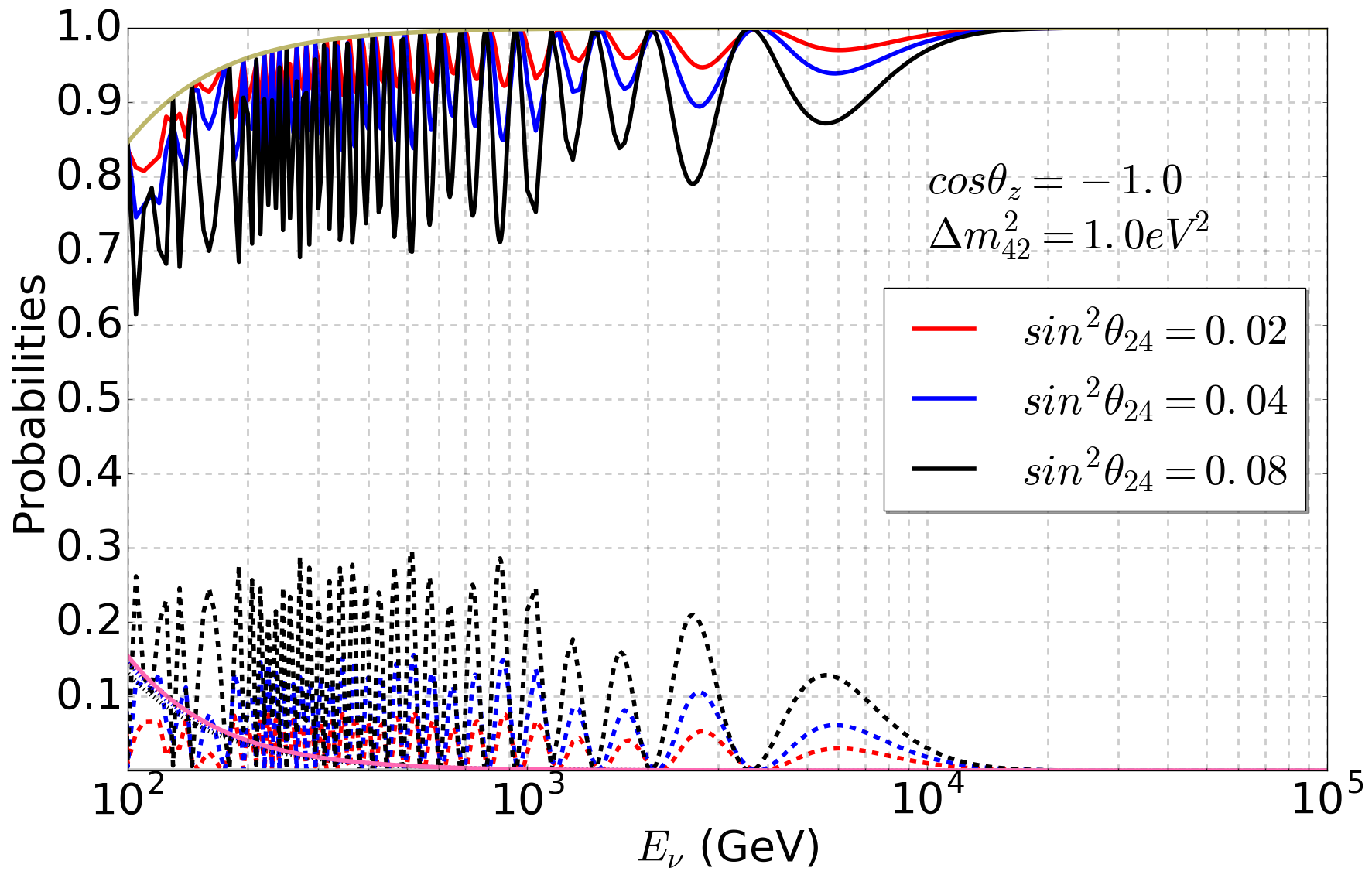}
\caption{\label{fig:prob_favor} Same as figure~\ref{fig:prob_mass} but for flavor mixing scheme probabilities.}
\end{figure}

Figures~\ref{fig:prob_mass} and \ref{fig:prob_favor} show the probabilities for the mass- and flavor-mixing schemes, respectively. These Probabilities have the following properties. Because the matter potential is negative, the MSW resonance effect occurs in the anti-neutrino channel. This is shown as a single dip/peak located at the resonance energy $E \propto \Delta m_{42}^2/2V_\mu$, a few TeV for $\sim 1$~eV scale sterile neutrino, for mantle crossing trajectories with $\cos \theta_z = -0.6$. The depth (height) of the MSW resonance dip (peak) is proportional to $\sin^2\alpha$ or $\sin^2\theta_{24}$ in case of mass-mixing or flavor-mixing schemes, respectively. In the mass-mixing scheme, ${\bar \nu}_\mu$ dominantly converts to ${\bar \nu}_s$ at the resonance energy, while in the flavor-mixing scheme ${\bar \nu}_\mu$ entirely converts to ${\bar \nu}_s$. When $|\cos \theta_z| < 0.82$, the trajectories cross mantle-core-mantle and this parametric resonance is more complex, as shown in the $\cos \theta_z = -1$ plots. At the parametric resonance ${\bar \nu}_\mu$ dominantly converts to ${\bar \nu}_\tau$ for the mass-mixing scheme with $\sin^2\alpha > 0.02$, while again ${\bar \nu}_\mu$ entirely converts to ${\bar \nu}_s$ in the flavor mixing scheme. 

For the neutrino channel there is no dip or peak due to sterile mixing, but for energies less than $\sim 0.5$ TeV, the $\nu_\mu$ survival probability $P_{\mu\mu}$ is diminished due to \textcolor{black}{increase} of  $P_{\mu s}$ or $P_{\mu\tau}$ transition probability (see figures~\ref{fig:prob_mass} and \ref{fig:prob_favor}). The effect of 2-3 mixture also becomes substantial at energies $E<0.5$~TeV. For probability calculations below $100$~GeV, please consult reference~\cite{Razzaque:2012tp}.

\section{Fluxes}
Atmospheric neutrinos are produced as decay products in hadronic showers resulting from collisions of cosmic rays with nuclei in the atmosphere. At low energies the muon and electron neutrinos come mainly as secondary products of the pion and kaon mesons (known as the conventional atmospheric neutrino flux). At higher energies an extra contribution to the total flux is relevant due to prompt decay of the charmed mesons ($D^{0},D^+,D_{S}^+,\Lambda_{C}^+$) created from collisions of cosmic rays, and  a crossover between the conventional and prompt atmospheric neutrino fluxes occurs between $10^5-10^6$ GeV. For this work we have
used Gaisser-Honda model of the conventional muon neutrino flux\textcolor{black}{~\cite{Barr:2004br,Honda:2006qj,Gaisser:2011cc} extended to PeV neutrinos ~\cite{Aartsen:2013eka}} and Enberg-Reno-Sarcevic model for the prompt contribution~\cite{Enberg:2008te}.

After propagation through the Earth, primary fluxes of atmospheric muon ($\phi_{\mu}^0$) and electron ($\phi_{e}^0$) neutrinos can be used to calculate the $\nu_\mu$ flux at the detector as
\begin{align}
\phi_{\mu}=\phi_{\mu}^0 P_{\mu \mu}+\phi_{e}^0P_{e \mu}\approx 
\phi_{\mu}^0 P_{\mu \mu} \,.
\end{align}
A similar equation holds for the anti-neutrinos. Here the last approximation follows from the facts that $\phi_{\mu}^0 \gg \phi_{e}^0$ and $P_{e \mu} \ll 1$. However, we take into consideration the contribution from the tau leptons, created by the $\nu_{\tau}N$ charge-current (CC) interactions. The tau leptons decay into muons with branching ratio $\epsilon \sim 0.18$ and the corresponding event is recorded as $\nu_{\mu}^{\rm CC}$. The $\nu_{\tau}$ flux at the detector equals $\phi_{\tau}\approx \phi_{\mu}^0 P_{\mu \tau}$, since, again, $\phi_{\mu}^0 \gg \phi_{e}^0$ and $P_{e\tau} \ll 1$. To be detected as a $\nu_{\mu}$ event in the same muon energy bin the $\nu_{\tau}$ energy needs to be $\approx 2.5$ times higher than the $\nu_{\mu}$ energy.

\begin{figure}[tbp]
\centering
\includegraphics[width=.75\textwidth]{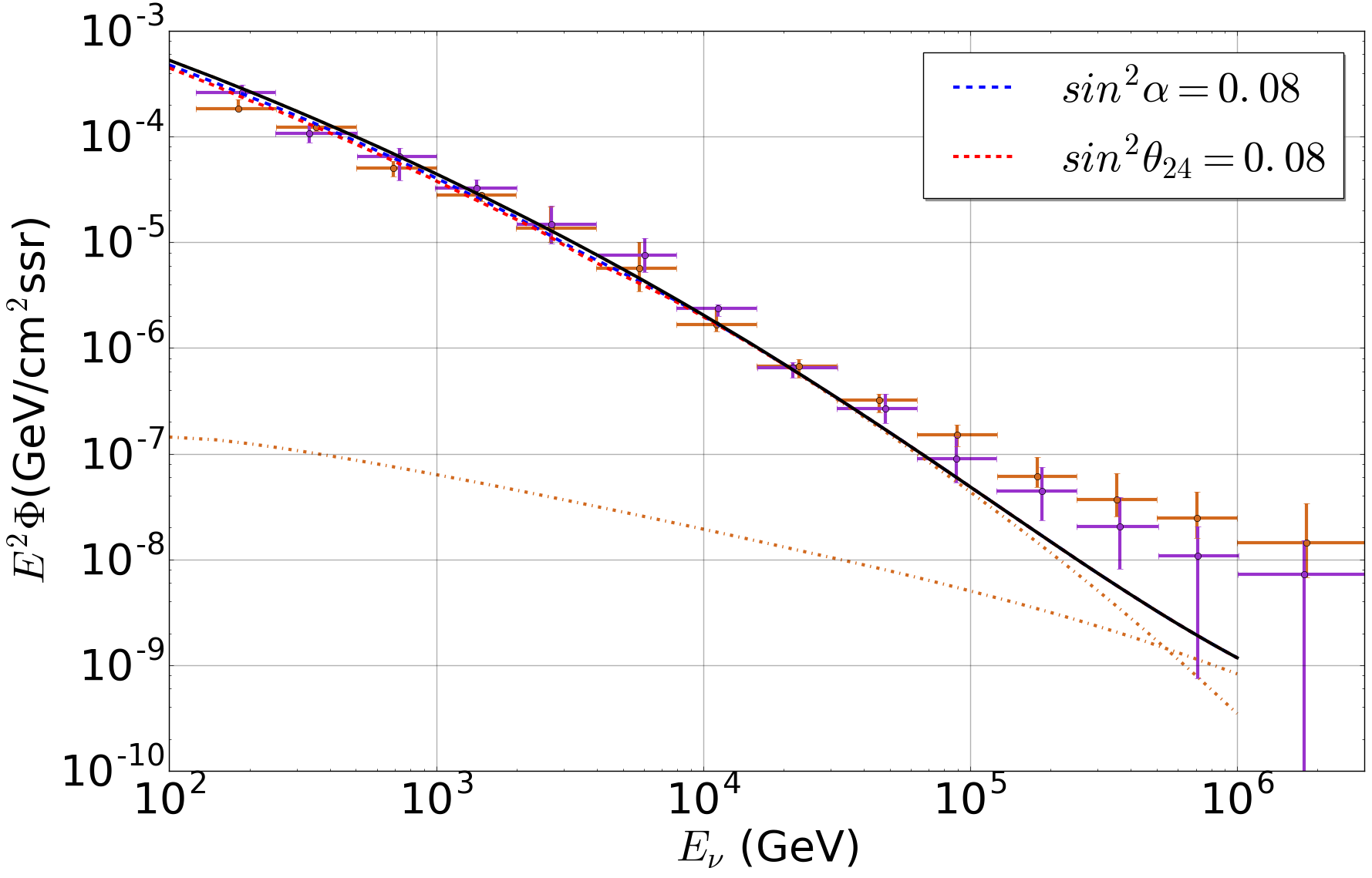}
\caption{\label{fig:flux} Zenith-angle ($86^\circ-180^\circ$) averaged atmospheric muon neutrino and antineutrino flux data~\cite{Aartsen:2017nbu} and models~\cite{, Enberg:2008te}.  The total flux is composed of conventional and prompt components (orange dot-dashed lines) and shown without $\nu_s$ mixing (black continuous line) and with $\nu_s$ mixing (blue-dashed line for mass mixing and red-dashed line for flavor-mixing).}
\end{figure}

In figure~\ref{fig:flux} we shown the initial muon neutrino and antineutrino flux, averaged in zenith-angle bin from $86^\circ$ to $180^\circ$. The data points are measurements by IceCube 79-string (magenta data points) and 86-string (orange data points) configurations~\cite{Aartsen:2017nbu}. The orange dot-dashed lines represent the conventional and prompt fluxes, while the black continuous line corresponds to the total. The effect of sterile-neutrino mixing is shown as the blue (red) dashed lines for the mass-mixing (flavor-mixing) scheme with $\sin^2\alpha=0.08$ ($\sin^2\theta_{24}=0.08$). The general feature of the sterile neutrino effect is a depletion of total flux in the $\approx 10^2-10^4$~GeV range. Note that the data points are systematically higher than the model for $E_\nu \gtrsim 10^5$~GeV. This is likely due to an astrophysical diffuse flux component. 

\section{IceCube data and event statistics}
We use publicly available IC86 data from the IceCube Collaboration's website\footnote{\url{http://icecube.wisc.edu/science/data/IC86-sterile-neutrino}} which has also been used in IceCube sterile neutrino study~\cite{TheIceCube:2016oqi}. This data release contains 20,145 well-reconstructed up going muon neutrino events, detected over a live time of 343.7 days (2011-2012). The data release also contains detector response arrays for the IC86 configuration and conventional atmospheric flux models. These tensors/arrays have the form: $T(E_{\nu}, \cos \theta_z, E_{\mu})$, where $E_{\mu}$ is the reconstructed muon energy (logarithmically spaced in 10 bins ranging from 400~GeV to 20~TeV), the muon direction is spaced in 21 bins from $\cos \theta_z = -1$ to 0.24, and the neutrino energy $E_\nu$ is logarithmically spaced in 200 bins from 200~GeV to $1 \times 10^6$~GeV. The tensors have units of GeV~cm$^2$~s~sr. 

\begin{figure}[tbp]
\centering
\includegraphics[width=.95\textwidth]{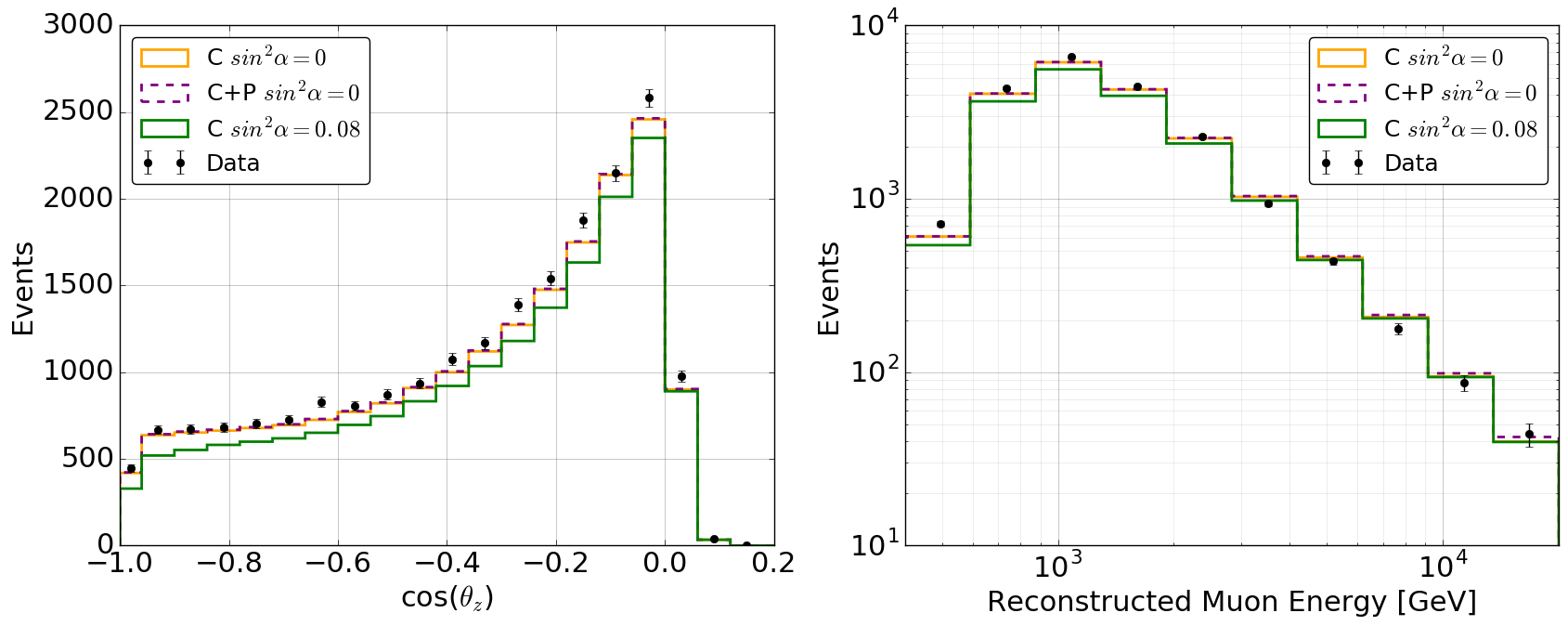}
\caption{\label{fig:events} Zenith angle (left panel) and neutrino energy (right panel) distributions of IC86 neutrino event data~\cite{TheIceCube:2016oqi} and models. The models shown are for conventional (C) atmospheric flux-only and without (with) $\nu_s$ mixing using orange (green) continuous histogram; and for conventional+prompt (C+P) atmospheric flux and without $\nu_s$ mixing using magenta dashed histogram.}
\end{figure}

We calculate the number of expected events with and without sterile neutrino mixing using the ``nominal'' detector response tensors that correspond to standard detector sensitivity according to  
\begin{align}
N(\cos\theta_{z}, E_{\mu})=\sum_{(E_{\nu})_{\textrm{bins}}} T(E_{\nu}, \cos\theta_{z}, E_{\mu}) \bar{\phi}(E_{\nu}, \cos\theta_{z})\,,
\label{eq:bins}
\end{align}
where $\bar{\phi}(\cos\theta_{z}, E_{\nu})$ is the flux at the detector, averaged for the same energy and zenith-angle bins as the tensor. These expected events can be compared with the experimental data. In figure~\ref{fig:events} we plot an example of zenith angle (left panel) and reconstructed muon energy (right panel) distributions of expected events based on given models and compare with IC86 neutrino event data~\cite{TheIceCube:2016oqi}. For this particular example we have used the $\nu_s$ mass-mixing scheme and atmospheric neutrino flux model with (histogram labeled as C+P) and without (histograms labeled as C) the prompt component. The effect of prompt flux component is evident at energies $\gtrsim 5\times 10^3$~GeV (right panel). The histograms labeled $\sin^2 \alpha=0$ correspond to no $\nu_s$ mixing and those models match with data better than the model with $\sin^2 \alpha=0.08$. We perform a detailed chi-square test next in order to derive constraint on the $\nu_s$ mixing parameters. 

\section{Statistical test}
In order to find sensitivity to sterile neutrino mixing parameters using experimental data with uncertainties in both model and data, we use the following $\chi^2$ function~\cite{Esmaili:2013vza}
\begin{align}
\nonumber
\chi^2(\sin^2 \theta^*, \Delta m_{42}^2; \hat{\beta})=
\sum_{i,j} \frac{[(N_{ij})_{\rm exp}-\beta_{0}\beta_{2}[1+\beta_{1}(0.56+(\cos\theta_{z})_{i})](N_{ij})_{\rm mod}]^2}{(\sigma_{ij}^2)_{\rm stat} +(\sigma_{ij}^2)_{\rm syst}}  \\ 
+ \frac{(1-\beta_{0})^2}{\gamma_{0}^2}+\frac{\beta_{1}^2}{\gamma_{1}^2}+\frac{(1-\beta_{2})^2}{\gamma_{2}^2}+\frac{\beta_{3}^2}{\gamma_{3}^2} \,,
\label{chi2}
\end{align}
where $\theta^* = \alpha$ or $\theta_{24}$ represents the mass mixing scheme or flavor
mixing scheme, respectively. The parameters $\hat{\beta} = (\beta_{0}, \beta_{1}, \beta_{2}, \beta_{3})$ take into account the uncertainties of the atmospheric neutrino flux normalization; zenith dependence tilt; muon to electron neutrino ratio; and power-law index of the conventional flux, respectively. Experimental data and theoretical model data are represented as $(N_{ij})_{\rm exp}$ and $(N_{ij})_{\rm mod}$, respectively, with the $i$-th ($j$-th) index refers to $\cos\theta_z$ ($E_\mu$) bin according to equation~(\ref{eq:bins}). The distribution of events can be rotated around the point $\cos \theta_z=-0.56$ (middle point in the $\cos\theta_{z} = [0,0.12]$ range of response tensors) with an angle determined by $\beta_{1}$. The errors on the fitting parameter set $\hat{\beta}$ used for calculation are: $\gamma_{0}=0.2$, $\gamma_{1}=0.04$, $\gamma_{2}=0.05$ and $\gamma_{3}=0.1$. The statistical error is calculated as $(\sigma_{ij})^2_{\rm stat} = (N_{ij})_{\rm exp}$ and the uncorrelated systematic error as $(\sigma_{ij})^2_{\rm syst}=f^2 (N^2_{ij})_{\rm exp}$ with a parameter $f$, which quantifies the detector precision. We will present constraints on sterile neutrino mixing parameters with $f=5\%$ and $10\%$.
 
We minimize the $\chi^2$ function in equation~(\ref{chi2}) by varying the $\hat{\beta}$ parameters for fixed values of the mixing angle $\sin^2 \theta^*$ and mass-squared difference $\Delta m_{42}^2$. The difference in minimized $\chi^2$ for models with and without $\nu_s$ mixing 
\begin{align}
\triangle \chi^2=\chi^2_{\rm min}(\textrm{with sterile neutrino})-\chi_{\rm min}^2(\textrm{no sterile neutrino})\,,
\end{align}
is then used to quantify the rejection significance of the $\nu_s$ mixing parameters for a given mixing scheme. In case of no $\nu_s$ mixing and for $f=10\%$, using both the conventional and prompt atmospheric fluxes we obtain $\chi_{\rm min}^2 = 178$ for 204 degrees of freedom.  The best-fit parameter values in this case are $\beta_0 = 1.01$, $\beta_1 = 0.0003$, $\beta_2 = 1.0006 $ and $\beta_3 = 0.02$. We report constraints on $\sin^2 \theta^*$ and $\Delta m_{42}^2$ derived from our analysis of IC86 data in the next section.

\section{Results and Discussion}

\begin{figure}[tbp]
\centering
\includegraphics[width=.45\textwidth]{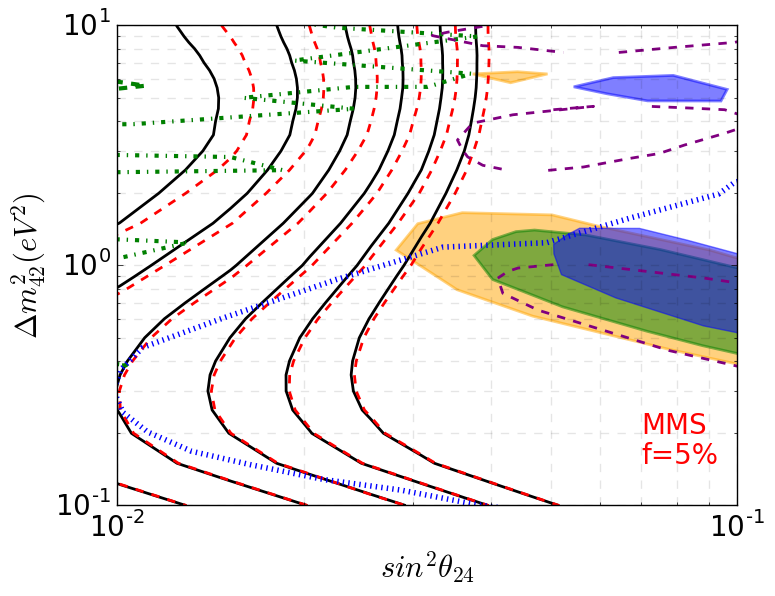}
\includegraphics[width=.45\textwidth]{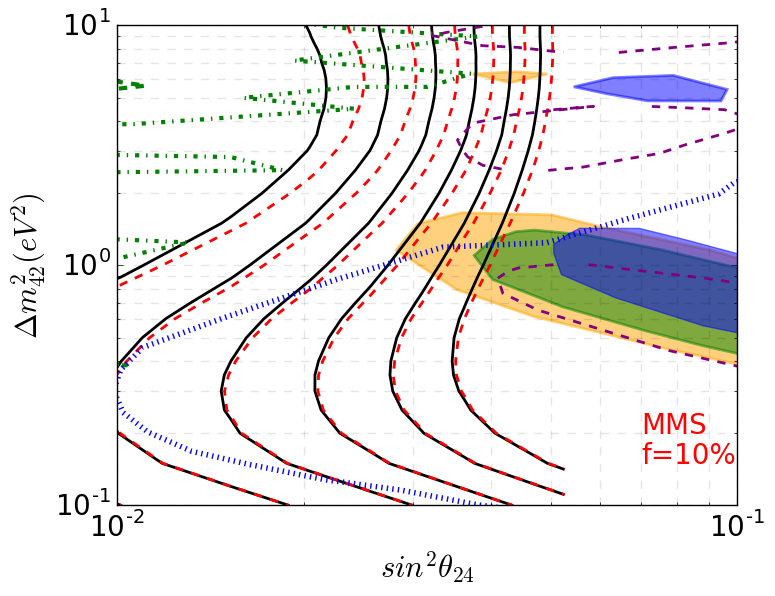}
\includegraphics[width=.45\textwidth]{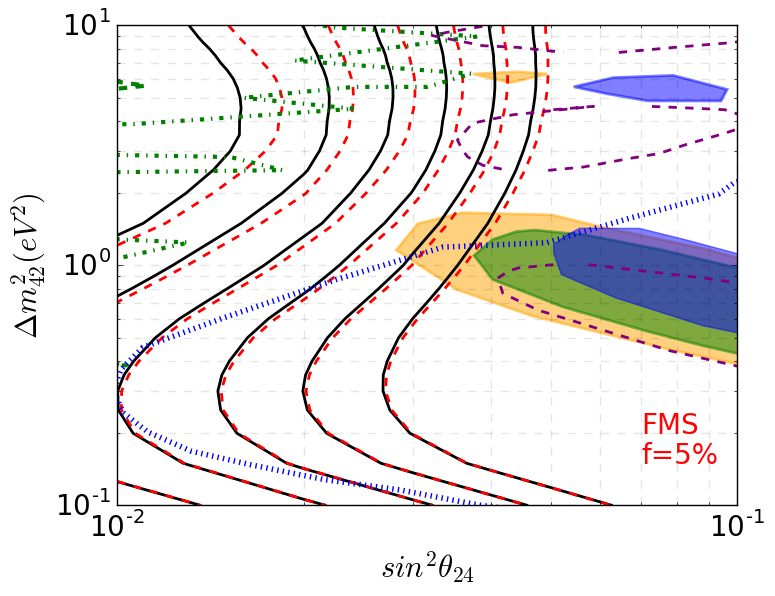}
\includegraphics[width=.45\textwidth]{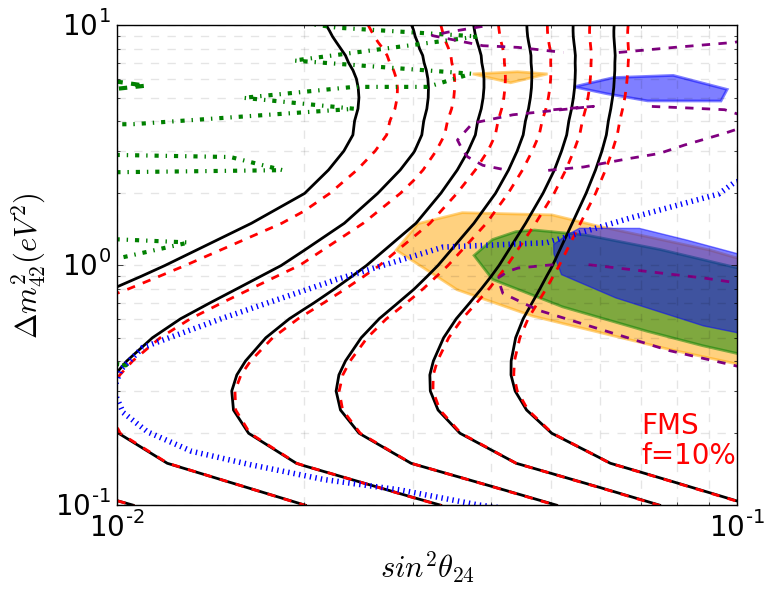}
\caption{\label{fig:exclusion} Exclusion regions in the sterile neutrino mixing parameter space for the mass-mixing (\textcolor{black}{MMS} at top panels) and flavor-mixing (\textcolor
{black}{FMS} at bottom panels) schemes. The left (right) panels are for f = 5\% (10\%) uncorrelated systematic uncertainties. The continuous black lines correspond to confidence levels from $1\sigma$ to $6\sigma$ with initial conventional atmospheric flux and the red dashed lines correspond to the same but including also the atmospheric prompt flux contribution. The blue dotted line corresponds to IceCube's 99\% CL exclusion region~\cite{TheIceCube:2016oqi} and the green dot-dashed line represents $99\%$ CL exclusion region from the combined disappearance experiments~\cite{Dentler:2018sju}. The shaded areas correspond to allowed regions from the appearance experiments. The latest MiniBooNE $3\sigma$ allowed regions are shown with purple dashed contours~\cite{Aguilar-Arevalo:2018gpe}. See main text for more details.}
\end{figure}

\subsection{Constraints on $\nu_s$ mixing parameters}
The results of our analysis of IC86 public data are shown in figure~\ref{fig:exclusion}, with the top two panels for the mass-mixing scheme, by converting the angle $\alpha$ to $\theta_{24}$ according to equation~(\ref{alpha_t24}), and the bottom two panels for the flavor-mixing scheme. 
The exclusion regions in the $\Delta m_{42}^2$--$\sin^2\theta_{24}$ parameter space are right to the black continuous curves at $1\sigma$ -- $6\sigma$ CL (from left to right) according to our analysis using the conventional atmospheric flux only. The red dashed lines represent the same but using both the conventional and prompt atmospheric fluxes.  The left (right) panels are for $f=5\%$ ($10\%$) uncorrelated systematic uncertainties. Larger systematics of course relax constraints on the mixing parameters. Furthermore, the flavor-mixing scheme also gives less tighter constraints on the allowed parameters. 
\textcolor{black}{This is because $2\sin^2 \theta _{24} \approx \sin^2 \alpha$, for small values of $\sin^2 \alpha$ according to equation (\ref{alpha_t24}), on each point of the exclusion curves in figure~\ref{fig:exclusion}; which makes the flavor-mixing scheme the less restrictive model.}

Also, in figure~\ref{fig:exclusion} we show IceCube $99\%$ CL exclusion region~\cite{TheIceCube:2016oqi} with the blue dotted line and a $99\%$ CL exclusion region using combined disappearance experiments, including IceCube, with the green dot-dashed line~\cite{Dentler:2018sju}. The orange shaded areas represent allowed regions with $99.73\%$ CL ($3\sigma$) for combined appearance experiments including decay-in-flight (DiF) data. The blue shaded areas are allowed regions at $99\%$ CL and excluding the DiF data, while the green shaded area is the same allowed region at $99\%$ CL but including the DiF data. The purple dashed contours for the MiniBooNE latest result correspond to $3\sigma$ allowed region~\cite{Aguilar-Arevalo:2018gpe}. When it is necessary to convert mixing angles from appearance or disappearance experiments we use the formula
\begin{align}
\sin^2\theta_{24} = \frac{\sin^2 2\theta_{\mu e}}{4|U_{e4}|^2(1-|U_{e4}|^2)} \,,
\end{align}
with $|U_{e4}| = 0.1$~\cite{Dentler:2018sju}.

We note that our exclusion regions begin to be less restrictive for $\Delta m^2_{42} \gtrsim 1$~eV$^2$, when taking into account the prompt atmospheric flux contribution. This change is greater 
as the value of the mass-square difference increases and is greater also for smaller sigma values. Although the difference between the significance regions with and without the prompt component 
is small for the range of energies studied here, our study suggests that prompt-type contribution could be important for $\Delta m_{42}^2 \gtrsim 10$~eV$^2$ and $\sin^2 \theta_{24} \lesssim 0.01$.
\textcolor{black}{This is because the prompt flux contribution increases the total number of events at higher energies which tends to replenish the lost events caused by sterile neutrino mixing, thus relaxing slightly the bound limits}. 
We find that the  $3\sigma$ allowed regions from the combined appearance experiments (orange shaded areas) are excluded at $\gtrsim 3\sigma$ ($\gtrsim 4\sigma$) CL in case of flavor-mixing 
(mass-mixing) scheme with $10\%$ systematics, both with and without the prompt atmospheric flux component. The $3\sigma$ allowed regions from the latest MiniBooNE appearance results 
are mostly excluded at $\gtrsim 3\sigma$ CL in case of the flavor-mixing scheme, when considering only the conventional atmospheric flux component; but some smaller regions survive when 
including the prompt flux component (see bottom right panel of figure~\ref{fig:exclusion}). In case of mass-mixing scheme, most allowed regions from the latest MiniBooNE appearance results 
are excluded at $\gtrsim 3\sigma$ CL, both with and without the prompt atmospheric flux component (see top right panel of figure~\ref{fig:exclusion}).\\


\subsection{Summary}
In this paper we have used the mass mixing scheme and $\nu_\mu - \nu_s$ flavor mixing scheme to calculate the effect of one sterile neutrino on oscillations through propagation inside the Earth. We have analyzed the cases with initial conventional and conventional + prompt atmospheric neutrino flux to see the differences of the two scenarios, for both schemes. In order to calculate the number of events and sensitivity of the IceCube neutrino detector to sterile neutrino mixing, we have used publicly available IceCube data tensors for detector characteristics in the muon energy range of 200--$10^4$ GeV.

We have performed a chi-square test of the sterile-neutrino mixing models using measured event distributions in energy and zenith-angle bins~\cite{TheIceCube:2016oqi} in order to calculate the exclusion regions in the $\Delta m_{42}^2$--$\sin^2\theta_{24}$ plane. We find that the flavor-mixing scheme in general provides less restrictive constraints on the mixing parameters, as expected. Our results are also in general more restrictive than IceCube limit, in particular for $\Delta m_{42}^2 \gtrsim 1$~eV$^2$. 

An important finding of our analysis is that including the prompt atmospheric flux contribution in the sterile neutrino analysis is required. This is because neutrinos with energy up to $10^6$~GeV is used for muon event calculation in IceCube data and the prompt component becomes significant at $\gtrsim 10^5$~GeV. The prompt component relaxes slightly the exclusion regions above $\Delta m^2 _{42} \sim 1$~eV$^2$. Future studies could give larger differences for $\Delta m_{42}^2 \gtrsim 10$~eV$^2$ and $\sin^2 \theta_{24} \lesssim 0.01$ values. In principle a cosmogenic flux component could be required as well when probing large $\Delta m^2 _{42}$ values. 
 
We exclude the combined appearance experiments' $3\sigma$ CL allowed regions in the $\Delta m_{42}^2$--$\sin^2\theta_{24}$ parameter space at $\gtrsim 3\sigma$ CL, in case of less restrictive flavor-mixing scheme and with $10\%$ uncorrelated systematics. The $3\sigma$ CL allowed regions from the latest MiniBooNE results are excluded at $\gtrsim 3\sigma$ CL when using conventional atmospheric flux only. Some small allowed regions survive at $\sim 3\sigma$ CL when we include prompt atmospheric flux component, thus keeping tension between the appearance and disappearance experiments.

\acknowledgments

We would like to thank Carlos Arg\"uelles, Darren Grant and Dawn Williams for helping us to understand IceCube public data release. We also thank Mary Hall Reno for kindly providing the prompt atmospheric flux model data. This work was supported by a National Research Foundation (South Africa) grant with Grant No: 111749 (CPRR).  



\end{document}